\documentclass[aps,twocolumn,superscriptaddress]{revtex4-2}

\usepackage{amsmath,charter,multirow}
\usepackage{graphicx,longtable,xfrac}
\usepackage[colorlinks=true,citecolor=blue,linkcolor=blue,breaklinks=true]{hyperref}
\usepackage{amssymb}
\usepackage{amsmath}
\usepackage{color}
\usepackage{bm}
\usepackage{soul}
\usepackage{indentfirst}

\begin{document}

\title{Hidden frustration in the triangular-lattice antiferromagnet NdCd$_3$P$_3$}

\author{Juan R. Chamorro}
\altaffiliation{Present address: Department of Materials Science and Engineering, Carnegie Mellon University, Pittsburgh, Pennsylvania 15213, USA}
\affiliation{Materials Department, University of California, Santa Barbara, California 93106, USA}

\author{Steven J. Gomez Alvarado}
\affiliation{Materials Department, University of California, Santa Barbara, California 93106, USA}

\author{Dibyata Rout}
\affiliation{Materials Department, University of California, Santa Barbara, California 93106, USA}

\author{Sarah Schwarz}
\affiliation{Materials Department, University of California, Santa Barbara, California 93106, USA}

\author{Allen Scheie}
\affiliation{MPA-Q, Los Alamos National Laboratory, Los Alamos, New Mexico 87545, USA}

\author{Ganesh Pokharel}
\affiliation{Materials Department, University of California, Santa Barbara, California 93106, USA}
\affiliation{Perry College of Mathematics, Computing, and Sciences, University of West Georgia, Carrollton, GA 30118, USA}

\author{Alexander I. Kolesnikov}
\affiliation{Neutron Scattering Division, Oak Ridge National Laboratory, Oak Ridge, Tennessee 37831, USA}

\author{Lukas Keller}
\affiliation{Laboratory for Neutron Scattering and Imaging, Paul Scherrer Institute, CH-5232 Villigen-PSI, Switzerland}

\author{Stephen D. Wilson}
\email[]{stephendwilson@ucsb.edu}
\affiliation{Materials Department, University of California, Santa Barbara, California 93106, USA}

\begin{abstract}
We report a study of the magnetic ground state and crystal electric field (CEF) scheme in the triangular-lattice antiferromagnet NdCd$_3$P$_3$. Combined neutron scattering, magnetization, and heat capacity measurements demonstrate that the Nd$^{3+}$ moments occupying the triangular lattice in this material harbor hidden signs of frustration not detected in typical Curie-Weiss-based parameterization of the frustration index ($f=\Theta_{CW} / T_N$).  This is evidenced by a zero-field splitting of the Kramers' ground state and first excited state doublets at temperatures far in excess of $T_N$ as well as signatures of low-energy fluctuations for $T>>T_N$. A suppression of the zero-field ordered moment relative to its field saturation value is observed, and the impacts of this magnetic frustration as well as the coexisting bond frustration in the CdP honeycomb network on the physical properties of NdCd$_3$P$_3$ are discussed. 
\end{abstract}

\maketitle

\section{Introduction}

Geometrically frustrated lattices give rise to complex electronic states that can form due to competing spin, orbital, and charge interactions. The electronic degeneracies created through geometric frustration can lead to enhanced quantum fluctuations that facilitate the formation of long-range entangled macrostates, such as the quantum spin liquid state \cite{Savary2016, Broholm2020}. Systems with structural motifs such as triangles (\textit{e.g.} kagome, triangular, honeycomb lattices) or tetrahedra (\textit{e.g.} pyrochlore, diamond lattices) often possess strong geometric frustration, which in some cases stabilizes exotic magnetic states \cite{Chamorro2020}.

The two-dimensional triangular-lattice, in particular, is the prototypical quantum spin liquid host when populated with antiferromagnetically-interacting \textit{S} = 1/2 magnetic moments \cite{Misguich1999}. A number of Yb-based triangular lattice systems, such as YbMgGaO$_4$ \cite{Li2015,Paddison2016} and NaYb\textit{X}$_2$ (\textit{X} = O, S, Se) \cite{Ding2019,Bordelon2019,Baenitz2018,Sarkar2019,Ranjith2019,Dai2021}, are now known to host various quantum disordered magnetic states owing to frustration between \textit{S}$_\mathrm{eff}$ = 1/2 Yb$^{3+}$ moments. 

\textit{R}Cd$_3$P$_3$ compounds possess a similar, ideal triangular net of lanthanide moments to YbMgGaO$_4$ and \textit{ARX}$_2$-type compounds and are predicted to be promising hosts for unusual, frustration-driven magnetic ground states \cite{Li2016}. While \textit{R}Cd$_3$P$_3$ is not stable with \textit{R} = Yb, a frustrated magnetic state is reported in lighter lanthanide members like CeCd$_3$P$_3$ \cite{PhysRevB.99.245159,Uzoh2023}. Little is known about the other variants, and, Nd$^{3+}$, in particular, is of interest due to its status of the lone remaining magnetic Kramers' ion in the reported \textit{R}Cd$_3$P$_3$ series.

Recently, the synthesis and bulk properties of the \textit{S}\textsubscript{eff} = 1/2 triangular-lattice antiferromagnet NdCd$_3$P$_3$ were reported \cite{Chamorro2023}, opening this material up for deeper exploration. Triangular-lattice layers of Nd$^{3+}$ are separated in excess of 10 $\mathrm{\AA}$ by cadmium phosphide layers which incorporate uncommon trigonal-planar CdP$_3$ units rarely observed in extended solids. These units exhibit bond frustration and short-range order \cite{arXiv2025LnCd3P3}, making the broader class of magnetic \textit{R}Cd$_3$P$_3$ compounds rare hosts of interleaved frustrated lattices (bond and magnetic). The magnetic layers are ideal two-dimensional triangular-lattices that are well-isolated from one another and yet potentially subject to local charge fluctuations due to the neighboring bond kagome-ice type bond frustration. 

A transition into a magnetically ordered ground state occurs in NdCd$_3$P$_3$ at \textit{T}\,=\,0.34 K \cite{Chamorro2023}. The depopulation of low-energy crystal electric fields (CEF) results in a low-temperature Curie-Weiss region between \textit{T} = 2 to 4.2 K with effective moment \textit{p}$_\mathrm{eff}$ = 1.98 $\mu_B$ and Weiss temperature $\Theta_{CW}$ = $-$0.38 K. This depopulation is confirmed in heat capacity measurements where a Schottky anomaly is observed with a maxima at \textit{T} = 18 K with a concomitant entropy release of Rln(2). These observations are consistent with a net \textit{S}\textsubscript{eff} = 1/2 ground state that is seemingly unfrustrated; however the details of the CEF scheme and ground state magnetic structure remain unexplored to date.

In this paper, we present neutron scattering, magnetization, and heat capacity data studying polycrystalline NdCd$_3$P$_3$. Inelastic neutron scattering (INS) measurements determine the ground state doublet and CEF level scheme, revealing an isolated \textit{S}$_\mathrm{eff}$ =1/2 ground state doublet. The magnetic structure of NdCd$_3$P$_3$ is determined to be \textit{A}-type antiferromagnetic order, with moments aligning ferromagnetically in-plane and antiferromagnetically in the interlayer direction, albeit with an ordered moment magnitude smaller than the field-polarized moment.  An anomalous splitting in the lowest energy CEF excitation is observed at temperatures far exceeding \textit{T}$_N$, demonstrating that the Kramers degeneracy is broken, which is also evidenced in heat capacity data by a zero-field splitting of the Nd$^{3+}$ Kramers ground state doublet. We propose this CEF level splitting and the reduced ordered moment signify strong frustration effects and an extended regime of fluctuations/short-range order not captured in a conventional Curie-Weiss analysis of the mean field ($\Theta_{CW}$) in NdCd$_3$P$_3$. 

\section{Experimental Details}

\subsection{Powder synthesis}
Non-isotopically enriched polycrystalline samples of NdCd$_3$P$_3$ were prepared using methods discussed in a previous report \cite{Chamorro2023}. Samples for neutron scattering measurements were prepared using isotopically enriched (isotope purity 99\%+) $^{114}$Cd purchased from Trace Sciences International to minimize neutron absorption. A modified synthesis route was required due to the elemental form of $^{114}$Cd, though this approach may lead to undesired secondary phases if not carefully controlled. $^{114}$Cd chunk was first pre-reacted with Nd in a 1:3 ratio within a crucible in a sealed, evacuated silica ampoule in order to minimize Cd vapor pressures at high temperature. The resulting brittle sample was then ground, mixed with phosphorus, and sealed in an evacuated silica ampoule. The reaction vessel was heated to 850 $^{\circ}$C for 48 hours and then quenched. The contents were removed, ground, resealed, and then retreated under the same conditions. All chemical handling was performed in an argon-filled glove box.

\subsection{Magnetization and heat capacity measurements}
Magnetic susceptibility measurements were carried out in a Quantum Design Magnetic Property Measurement System (MPMS3), as well as using a vibrating sample mode (VSM) option on a Quantum Design 14 T Dynacool Physical Property Measurement System (PPMS). Heat capacity measurements were performed using the PPMS and its measurement option, and measurements between \textit{T} = 0.1 and 2 K were performed using a dilution refrigerator insert for the PPMS.

\subsection{Neutron Scattering}

INS measurements at \textit{T} = 6 K were performed using the SEQUOIA spectrometer at the Spallation Neutron Source at Oak Ridge National Laboratory (ORNL) \cite{Granroth2010}.  Incident energies of $E_i=11$ and $E_i=55$ meV were used, selected by the high-resolution and high-flux Fermi choppers, respectively. A 4 g powder sample of Nd$^{114}$Cd$_3$P$_3$ was loaded and sealed into a 0.25" diameter aluminum can in a helium-filled glove box. The can with the sample was attached to the cold-head of a bottom-loading closed-cycle $^4$He refrigerator. The background contribution from the Al can was subtracted from the data by obtaining an empty can measurement at each temperature and incident energy. INS data were fit to point-charge models from structural data and independently to CEF fits using \textsc{PyCrystalField} \cite{Scheie2021}.

Powder neutron diffraction measurements were performed at temperatures between \textit{T} = 0.07 and 0.7 K using a dilution refrigerator at the DMC instrument \cite{Schefer1990} at the Swiss Spallation Neutron Source SINQ with an incident wavelength of $\lambda$ = 2.453 $\mathrm{\AA}$. A nuclear and magnetic structural model was refined against the data
using FullProf \cite{rodriguez2001fullprof} with symmetry mode analysis via SARAh \cite{Wills2000}. Sample powder was loaded in a copper can and copper diffraction peaks were co-refined in the structural analysis. 

\section{Results}

\subsection{Magnetization}

\begin{figure}[t]
    \includegraphics[width=0.45\textwidth]{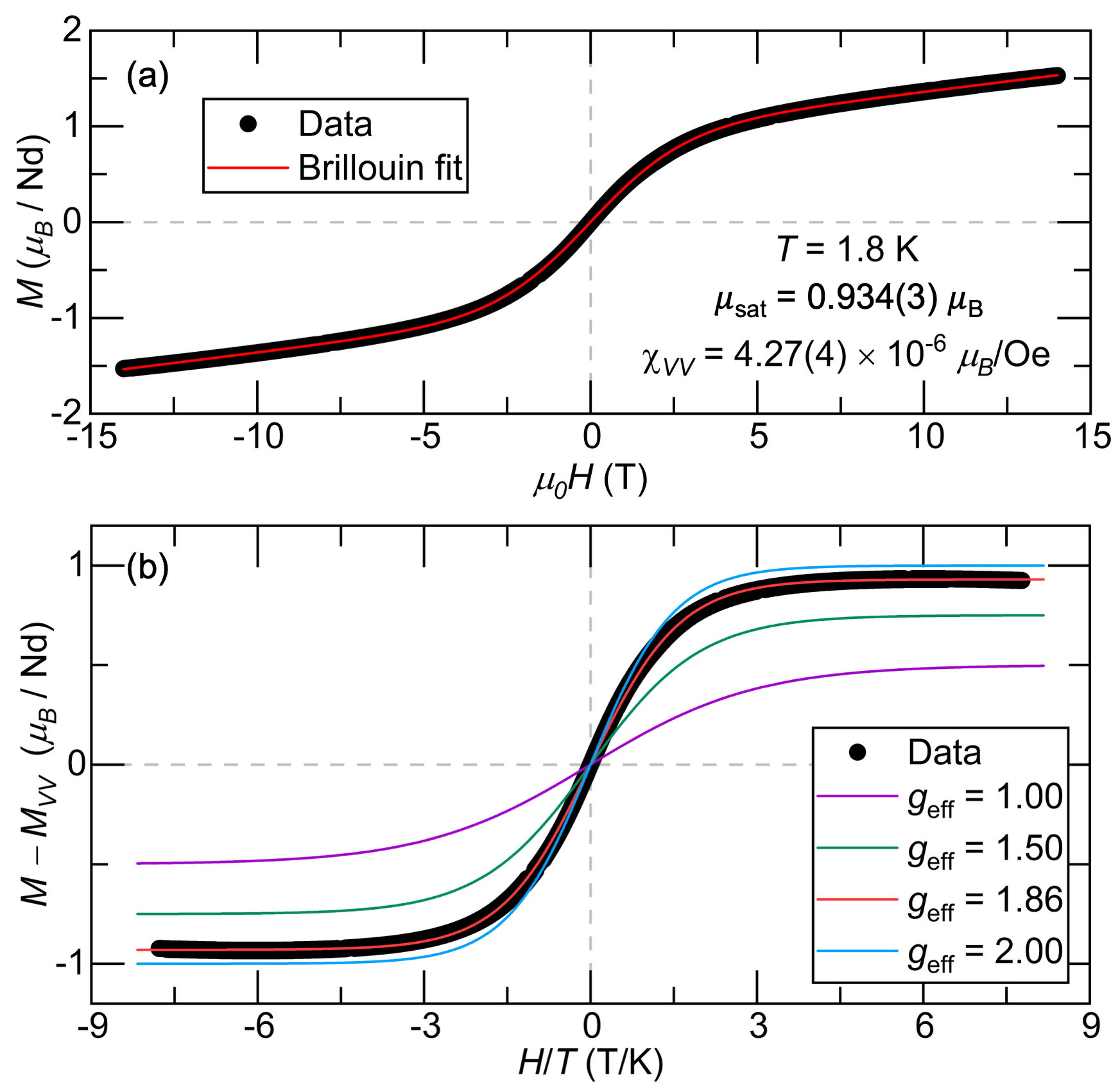}
    \caption{
    (a) Field-dependent magnetization $M(H)$ at $T = 1.8$ K with spin-$\frac{1}{2}$ Brillouin + Van Vleck fit. (b) Scaled plot of $M$ vs $H/T$ after Van Vleck subtraction, showing agreement with $g_\mathrm{eff} = 1.86$.}
\end{figure}

To determine the saturated local moment and powder-averaged g-factor, isothermal magnetization measurements were collected on polycrystalline NdCd$_3$P$_3$ at \textit{T} = 1.8 K and up to $\mu_0 H = \pm$ 14 T. At this low temperature, only the ground-state Kramers doublet is expected to be populated (as verified later via INS measurements), and the magnetization is modeled within a \textit{S}$_{\mathrm{eff}}$ = 1/2 framework.  Figure 1(a) shows an isothermal magnetization curve at \textit{T} = 1.8 K along with a fit using a J=$\frac{1}{2}$ Brillouin function with an added Van Vleck (\textit{VV}) term. The model assumes a saturation moment \( \mu_{\mathrm{sat}} = \tfrac{1}{2} g_{\mathrm{eff}} \mu_\mathrm{B} \) and includes a field-independent susceptibility contribution $\chi_{VV}$:

\begin{equation*}
M(H) = N_A \cdot \left( \frac{g_{\mathrm{eff}} \mu_\mathrm{B}}{2} \right) \cdot B_{1/2}\left( \frac{g_{\mathrm{eff}} \mu_\mathrm{B} H}{k_B T} \right) + \chi_{VV} H
\end{equation*}

where \( B_{1/2}(x) \) is the Brillouin function for \( J = \tfrac{1}{2} \). Fitting the measured curve yields \textit{g}$_\mathrm{eff}$ = 1.869(7), $\mu_\mathrm{eff}$ = 1.618(6) $\mu_\mathrm{B}$, $\mu_\mathrm{sat}$ = 0.934(3) $\mu_\mathrm{B}$, and $\chi_{VV}$ = 4.27(4) $\times 10^{-6}$ $\mu_\mathrm{B}$/Oe.  

Figure 1(b) examines the same magnetization data in a scaled form, plotting \textit{H}/\textit{T} vs. \textit{M} after subtracting the fit Van Vleck term. In this representation, a J=$\frac{1}{2}$ magnetic system with a fixed \textit{g}-factor should collapse onto a universal Brillouin curve. As shown in the figure, the experimental data can be well-modeled with \textit{g}$_\mathrm{eff}$ = 1.86, while curves for \textit{g}$_\mathrm{eff}$ = 1.0, 1.5, and 2.0 deviate systematically. This modeling provides independent confirmation for the \textit{g}-factor obtained by direct fit in Figure 1(a) and the result is consistent with the earlier g$_\mathrm{eff}$ inferred from the Curie-Weiss-derived $\mu_\mathrm{eff}$  \cite{Chamorro2023}.  These data provide useful benchmarks for comparison with the antiferromagnetic ordered moment and \textit{g}-tensor determined from neutron scattering data. 

\subsection{Neutron diffraction}

\begin{figure}[t]
    \includegraphics[width=0.5\textwidth]{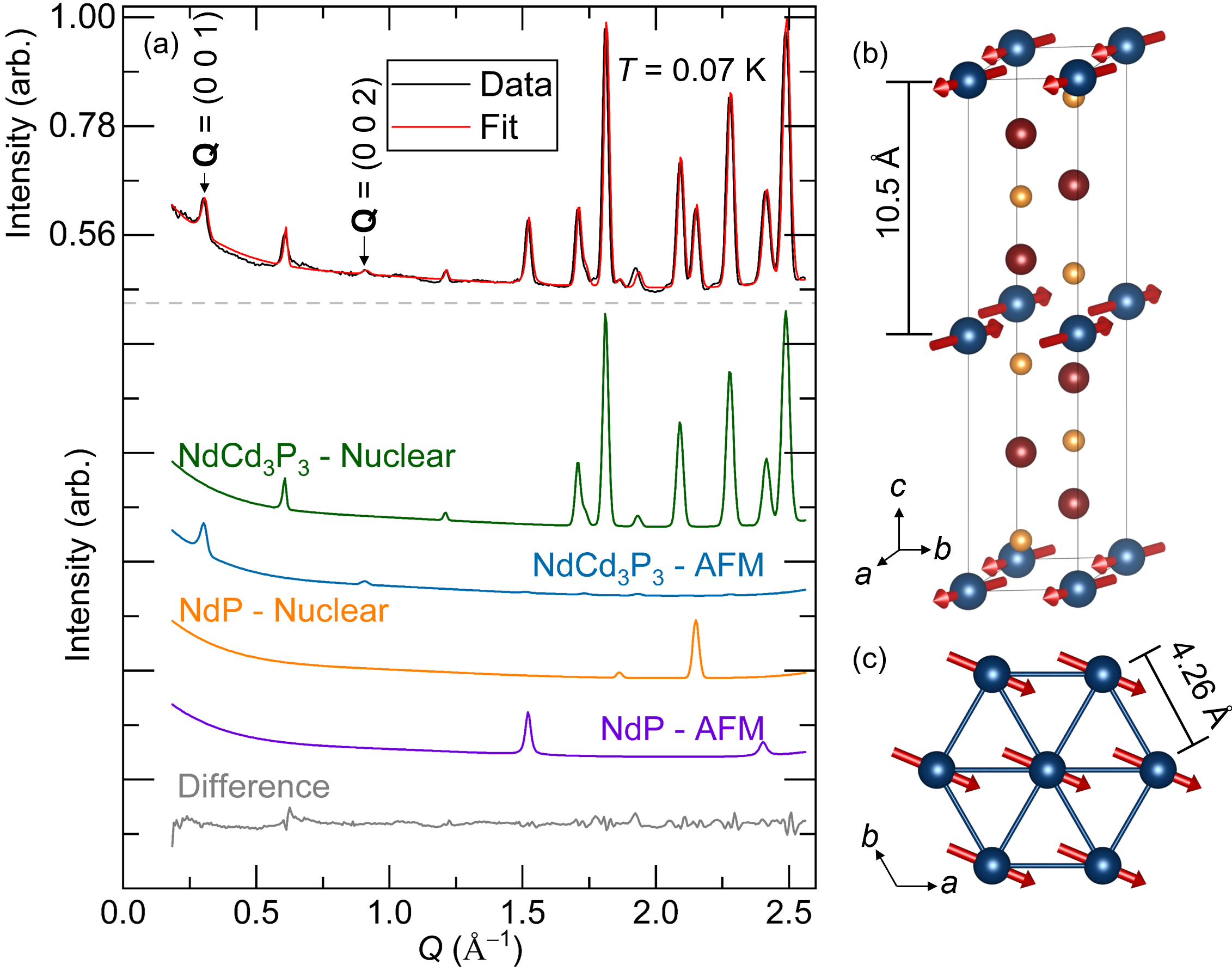}
    \caption{(a) Neutron diffraction pattern at \textit{T} = 0.07 K showing the overall Rietveld fit along with individual contributions from nuclear and magnetic Bragg peaks for NdCd$_3$P$_3$, as well as impurity nuclear and magnetic peaks from NdP. Magnetic peaks at $\mathbf{Q}$ = (0, 0, 1) and (0, 0, 2) are consistent with \textit{A}-type antiferromagnetic order. (b) Crystal structure of NdCd$_3$P$_3$ with $\mathbf{k}$ = (0, 0, 0) antiferromagnetic order. (c) Top-down view of a triangular layer of Nd$^{3+}$ demonstrating ferromagnetic coupling of moments in-plane.}
\end{figure}

In order to define the magnetic ground state of NdCd$_3$P$_3$, neutron diffraction measurements were performed.  Data collected at a base temperature of \textit{T} = 0.07 K are shown in Figure 2(a). Aside from the main NdCd$_3$P$_3$ phase, a 4.8(4)\%  (weight percent) NdP impurity phase is resolved. NdP is a known antiferromagnet below \textit{T}$_N$ = 11 K \cite{SchobingerPapamantellos1973}, and both its nuclear and magnetic structure were modeled within the powder pattern. Furthermore, while the magnetic peaks originating from NdP are stronger in intensity than those from NdCd$_3$P$_3$, they arise at well-isolated, distinct \textit{Q} values and can therefore be readily disentangled from the magnetic NdCd$_3$P$_3$ peaks. It is noted that contributions from NdP are not observed in other measurements, \textit{i.e.}, no indications of a long-range magnetic phase transition at \textit{T} = 11 K are observed in heat capacity; no extra crystal field excitations are observed in INS data; and including extra contributions to the magnetization do not result in statistically better fits.

The main NdCd$_3$P$_3$ phase was refined in the average, hexagonal \textit{P}6$_3$/\textit{mmc} structure, where Nd$^{3+}$ ions possess \textit{D}$_{3d}$ site symmetry, consistent with our previous report \cite{Chamorro2023}. Despite a complex local structure owing to charge instabilities in the cadmium phosphide blocking layers \cite{arXiv2025LnCd3P3}, the average long-range structure can be well-described with the hexagonal cell. Within this cell, the Nd$^{3+}$ triangular-lattice layers are separated from each other by 10.5 $\mathrm{\AA}$, as shown in Figure 2(b). The Nd$-$Nd cation distances within a given layer are 4.26 $\mathrm{\AA}$ apart and form an ideal triangular lattice. 

Data collected below \textit{T} = 0.34 K reveal new reflections consistent with long-range antiferromagnetic order. Two peaks can be observed in the data corresponding to \textbf{Q} = (0, 0, 1) and \textbf{Q} = (0, 0, 2), as highlighted in Figure 2(a). These reflections were indexed to a commensurate ordering wave vector of \textbf{k} = (0, 0, 0), corresponding to an \textit{A}-type antiferromagnetic order, where moments are aligned ferromagnetically within each layer and alternate antiferromagnetically along \textit{c}. This solution was identified by fitting multiple possible magnetic ordering models allowed by representational theory of the \textbf{k} = (0, 0, 0) structure, and the final fit yielded a goodness-of-fit one standard deviation lower than other attempted models. This antiferromagnetic order is shown schematically in Figure 1(b) and Figure 1(c), which illustrate the ferromagnetic intraplane and antiferromagnetic interplane correlations.  Moments align along the Cartesian axes (0.60(5), $-$0.38(4), 0.0)=($a$, $b$, $c$), \textit{i.e.}, within error along the (1, -1, 0) direction of the unit cell. The ordered moment was determined to be $\langle \mu \rangle$ = 0.71(5) $\mu_\mathrm{B}$, which is notably smaller than the apparent saturation magnetization value, $\mu_\mathrm{sat}$ = 0.934(3) $\mu_\mathrm{B}$ observed at \textit{T} = 1.8 K.  This implies the presence of substantial fluctuation effects or short-range order in the magnetic ground state of this compound.

\subsection{Inelastic neutron scattering and crystal field analysis}

\begin{figure*}[t]
    \includegraphics[width=0.75\textwidth]{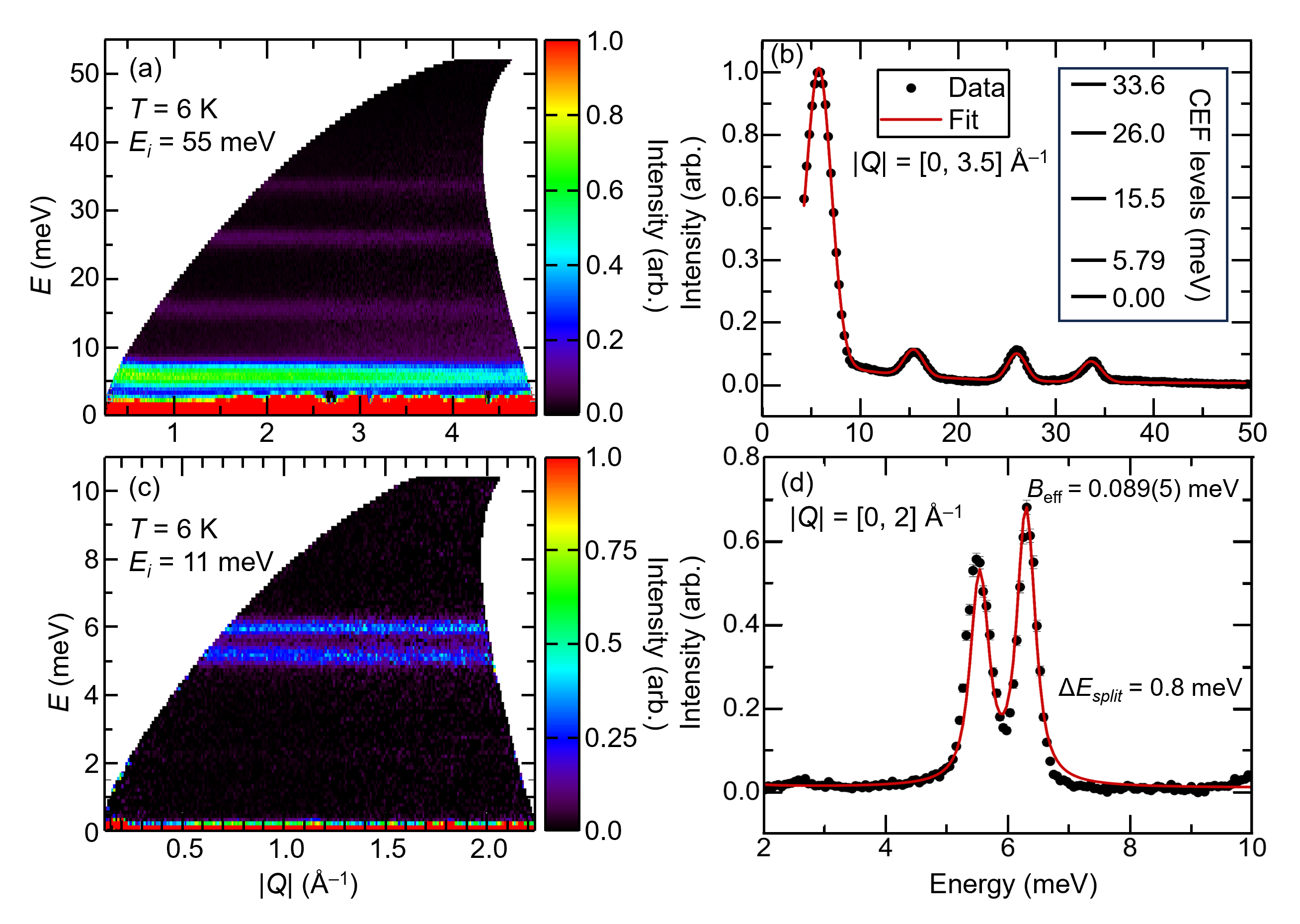}
    \caption{
    (a) Inelastic neutron scattering intensity $I(Q, \hbar\omega)$ at $T = 6$ K with $E_i = 55$ meV, showing CEF excitations of Nd$^{3+}$. (b) Momentum-integrated energy cut with CEF fit (solid line); inset: energy-level diagram of the five Kramers doublets. (c) High-resolution INS data with $E_i = 11$ meV resolving the first excited doublet. (d) Energy cut of the first excitation showing a $\Delta$\textit{E}$_{split}$ = 0.8 meV splitting; solid line shows a model including a static molecular field of \textit{B}$_\mathrm{eff}$ = 0.089(5) meV (root-mean-square magnitude determined from [0.11 (1), -0.11(1), 0.0] meV).}
\end{figure*}
\begin{table*} []
\caption{Eigenvectors and eigenvalues of CEF fit to the \textit{E}$_i$ = 55 meV INS data}
\begin{ruledtabular}
\begin{tabular}{c|cccccccccc}
E (meV) &$| -\frac{9}{2}\rangle$ & $| -\frac{7}{2}\rangle$ & $| -\frac{5}{2}\rangle$ & $| -\frac{3}{2}\rangle$ & $| -\frac{1}{2}\rangle$ & $| \frac{1}{2}\rangle$ & $| \frac{3}{2}\rangle$ & $| \frac{5}{2}\rangle$ & $| \frac{7}{2}\rangle$ & $| \frac{9}{2}\rangle$ \tabularnewline
 \hline 
0.0 & 0.0 & 0.0 & 0.304(7) & 0.0 & 0.0 & 0.941(3) & 0.0 & 0.0 & -0.147(5) & 0.0 \tabularnewline
0.0 & 0.0 & 0.147(5) & 0.0 & 0.0 & 0.941(3) & 0.0 & 0.0 & -0.304(7) & 0.0 & 0.0 \tabularnewline
5.787(5) & 0.00(8) & 0.0 & 0.0 & 0.5(2) & 0.0 & 0.0 & -0.84(13) & 0.0 & 0.0 & -0.17(5) \tabularnewline
5.787(5) & 0.17(5) & 0.0 & 0.0 & -0.84(13) & 0.0 & 0.0 & -0.5(2) & 0.0 & 0.0 & 0.00(8) \tabularnewline
15.47(5) & 0.0 & 0.263(14) & 0.0 & 0.0 & -0.334(8) & 0.0 & 0.0 & -0.905(5) & 0.0 & 0.0 \tabularnewline
15.47(5) & 0.0 & 0.0 & 0.905(5) & 0.0 & 0.0 & -0.334(8) & 0.0 & 0.0 & -0.263(14) & 0.0 \tabularnewline
26.03(3) & 0.0 & 0.0 & -0.297(14) & 0.0 & 0.0 & -0.052(7) & 0.0 & 0.0 & -0.954(4) & 0.0 \tabularnewline
26.03(3) & 0.0 & 0.954(4) & 0.0 & 0.0 & -0.052(7) & 0.0 & 0.0 & 0.297(14) & 0.0 & 0.0 \tabularnewline
33.63(6) & 0.0(3) & 0.0 & 0.0 & 0.09(7) & 0.0 & 0.0 & -0.14(5) & 0.0 & 0.0 & 0.99(13) \tabularnewline
33.63(6) & -0.99(13) & 0.0 & 0.0 & -0.14(5) & 0.0 & 0.0 & -0.09(7) & 0.0 & 0.0 & 0.0(3) \tabularnewline
\end{tabular}\end{ruledtabular}
\label{flo:Eigenvectors}
\end{table*}

To better understand the origin of the moment, INS measurements were performed at \textit{T} = 6 K to determine the CEF spectrum of Nd$^{3+}$ (\textit{J} = 9/2) in NdCd$_3$P$_3$. 2D intensity plots of the neutron scattering intensity \textit{I}(\textit{Q},\textit{E}) and their momentum-integrated energy cuts are shown in Figure 3. The Nd$^{3+}$ ions, situated in a local \textit{D}$_{3d}$ environment, are expected to exhibit five Kramers doublets arising from the splitting of the \textit{J} = 9/2 multiplet, which should generate four excited-state doublets in the neutron spectrum.

With an incident energy of $E_i=55$ meV, four sharp, well-defined excitations at \textit{E}$_1$ = 5.79, \textit{E}$_2$ = 15.5, \textit{E}$_3$ = 26.0, and \textit{E}$_4$ = 33.6 meV are observed, as shown in Figures 3(a,b). Previous heat capacity measurements detected a Schottky anomaly consistent with either a two-level gap of $\Delta$ = 6.4 meV or a three-level model with gaps of $\Delta_1$ = 4.3 and $\Delta_2$ = 8.1 meV \cite{Chamorro2023}. The INS data presented here support the two-level gap model, identifying the first, low-lying excited state at 5.79 meV.

To model the Nd$^{3+}$ CEF excitations, a Hamiltonian appropriate for \textit{D}$_{3d}$ site-symmetry was used:

\begin{equation*}
\mathcal{H}_{\mathrm{CEF}} = 
B_2^0\, \hat{O}_2^0 + 
B_4^0\, \hat{O}_4^0 + 
B_4^3\, \hat{O}_4^3 + 
B_6^0\, \hat{O}_6^0 + 
B_6^3\, \hat{O}_6^3 + 
B_6^6\, \hat{O}_6^6
\end{equation*}

where \textit{B}$^q_k$ are the CEF parameters and \( \hat{O}_k^q \) are Stevens operators. Starting from a point-charge model, the refined CEF parameters were determined via simultaneous fits to the INS spectrum and magnetization data. Uncertainty was estimated by the method in Ref \cite{Scheie2022}.  The best fit, shown in Figure 3(b), yielded the following crystal field parameters: \textit{B}$^0_2$ = 0.517 (3), \textit{B}$^0_4$ = $-$1.29(4) $\times 10^{-3}$, \textit{B}$^3_4$ = 9(2) $\times 10^{-3}$, \textit{B}$^0_6$ = 1(2) $\times 10^{-6}$, \textit{B}$^3_6$ = $-$1.41(4) $\times 10^{-3}$, and \textit{B}$^6_6$ = 7.3(3) $\times 10^{-4}$ meV. These parameters reproduce the observed energy spectrum and intensities, and determine a ground state Kramers doublet composed primarily of \(|\pm\tfrac{1}{2}\rangle\) character with minority components of \(|\pm\tfrac{5}{2}\rangle\) and \(|\pm\tfrac{7}{2}\rangle\) states. Table I contains a full list of the eigenvalues and eigenvectors for the CEF level scheme.

The best fit, with a reduced $\chi^2$ = 24.8, yields a ground-state doublet with XY anisotropy, though nearly comparable fits can be generated with Ising anisotropy, with a reduced $\chi^2$ = 32.2. The extracted \textit{g}-tensor components for the fit shown are \textit{g}$_{xx}$ = \textit{g}$_{yy}$ = \textit{g}$_{\perp}$ = 3.481(6) and \textit{g}$_{zz}$ = \textit{g}$_{\|}$ = 0.416(11), indicating that the magnetic moments are most easily polarizable in the plane perpendicular to the crystallographic \textit{c}-axis and consistent with the spin structure determined via neutron diffraction.  The ground-state doublet wavefunction is composed of: \(|\psi_{0,\pm}\rangle = 0.941(3)\,|\pm\tfrac{1}{2}\rangle \pm 0.304(7)\,|{\pm}\tfrac{5}{2}\rangle \pm 0.147(5)\,|\pm\tfrac{7}{2}\rangle\). As we will discuss later, this solution is constrained by the average symmetry of the lattice, which is known to be locally violated \cite{arXiv2025LnCd3P3}.  The inclusion of additional Stevens parameters in a lower symmetry point group will necessarily modify the wave function mixtures and resulting \textit{g}-tensor.

Added terms are undoubtedly necessary as the CEF-predicted, powder-averaged $g$-factor $g_\mathrm{avg,CEF} = \sqrt{\tfrac{1}{3}g_\parallel^2 + \tfrac{2}{3}g_\perp^2} = \sqrt{8.13} \approx$ 2.85 exceeds the $g_\mathrm{avg}$=1.86 determined via magnetization data. Based on this value, a $\mu_\mathrm{eff,CEF}$ = (\textit{g}$\sqrt{S(S+1)}$) $\mu_\mathrm{B}$ = 2.47 $\mu_\mathrm{B}$ and $\mu_\mathrm{sat,CEF}$ = (\textit{g}$_\mathrm{CEF}$/2) $\mu_\mathrm{B}$ = 1.43 $\mu_\mathrm{B}$ are expected. It is notable that the saturation moment in our magnetization data, even after the Van Vleck subtraction, reaches only $\mu_\mathrm{sat}$ = 0.934(3) $\mu_\mathrm{B}$. One possible explanation for this is preferential grain orientation creating a bias for $H\parallel c$, where NdCd$_3$P$_3$ crystallites are platelets with their faces normal to the crystallographic \textit{c}-axis. To test this conjecture, higher magnetic fields are required; however, our earlier Curie-Weiss analysis under low fields similarly found a local moment more consistent with the isothermal magnetization data \cite{Chamorro2023}.  The discrepancy between the maximum potential ordered moment from the CEF doublet and that seen by magnetization data may instead imply added terms in the CEF Hamiltonian that are currently unaccounted for.

Regardless of these added terms in the CEF ground state solution, INS data collected at 6 K with a lower incident energy show that the first excited state doublet is split with an energy of $\Delta$\textit{E}$_\mathrm{split}$ = 0.8 meV (Figure 3(c)). This is unexpected as the Nd$^{3+}$ CEF doublets are protected by time-reversal symmetry, and INS measurements were collected at temperatures greater than 17 $\times$ \textit{T}$_N$, thus static internal fields are not expected. In order to understand how a mean field can generate this splitting, various orientations for a molecular field along different crystallographic orientations were modeled. The best fit was achieved with a static field oriented along the (1, -1, 0) direction parallel to the in-plane moment orientation as determined in neutron diffraction measurements. The molecular mean field was modeled by modifying the Hamiltonian via: $$
\mathcal{H} = \mathcal{H}_{CEF} + \mu_B \bf{B} \cdot \bf{J}
$$
where $\bf{B}$ is the molecular field and $\bf{J}$ is a vector of quantum spin operators in the basis of the CEF Hamiltonian.

The mean field qualitatively reproduces the doublet splitting using a [0.11 (1), -0.11(1), 0.0] meV field parameterized using \textit{g} = 2, with a reduced $\chi^2$ = 11.59. This fit is shown in Figure 3(d) for data integrated over $|Q|$ = $[$0, 2$]$ \(\mathrm{\AA}^{-1}\). The presence of a strong mean field above $T_N$ implies that the ground state doublet should also show signs of splitting as well, though that splitting is expected to be below the energy resolution of our INS measurements.  Instead, to identify this potential ground state splitting, low-temperature heat capacity measurements were performed.

\subsection{Heat capacity}

\begin{figure}[t]
    \includegraphics[width=0.5\textwidth]{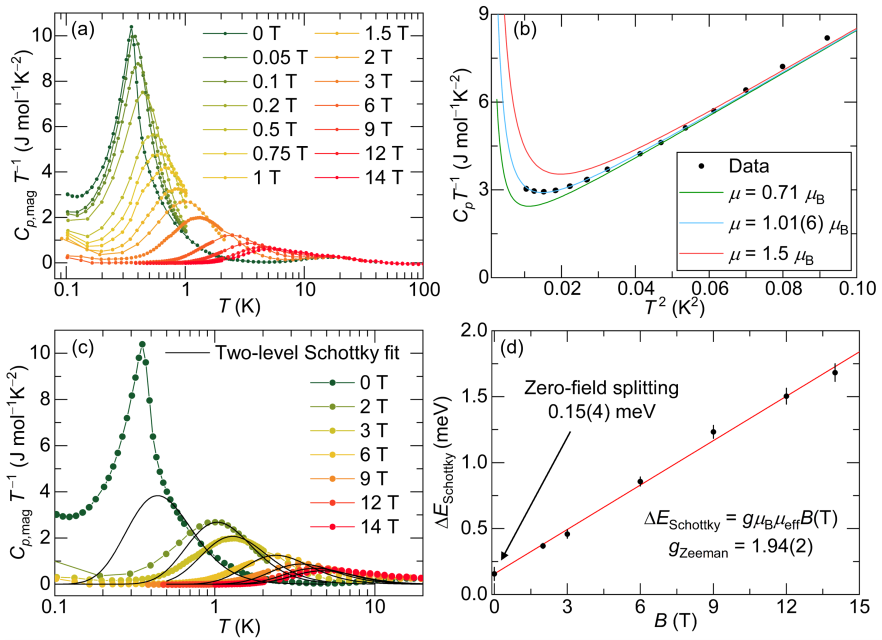}
    \caption{
    (a) Magnetic contribution to heat capacity \textit{C}$_{p,\mathrm{mag}}$\textit{T}$^{-1}$ from $\mu_0$\textit{H} = 0 to 14 T, showing the suppression and broadening of the $T_N$ anomaly and a broad Schottky tail. (b) Fits to a two-level Schottky model of the Schottky tail at selected fields. (c) Extracted Schottky gap $\Delta$\textit{E}$_\mathrm{Schottky}$ vs. applied field, showing linear Zeeman behavior with $g = 1.94(2)$.}
\end{figure}

Heat capacity measurements were performed down to \textit{T} = 0.1 K under a variety of magnetic fields. The magnetic contribution to the heat capacity was determined by subtracting a scaled phonon contribution using LaCd$_3$P$_3$ as a diamagnetic analog. Under zero-field, a peak indicative of magnetic order is observed at \textit{T} = 0.34 K, and, upon increasing the applied field, the transition both broadens and shifts to higher temperatures, as demonstrated in Figure 4(a). The data sets converge above \textit{T} = 18 K, where a broad, weakly field-dependent Schottky anomaly is centered. This is consistent with the two-level system associated between the Kramers doublet ground state and first excited CEF doublet, as discussed in the previous section and as fit previously \cite{Chamorro2023}.

The lowest data points in the zero-field \textit{C}$_p$(\textit{T}) data show an upturn from the Nd nuclear Schottky anomaly. Nd has two stable isotopes with non-zero nuclear spin, $^{143}$Nd and $^{145}$Nd, which both have \textit{I} = 7/2 and together make up 20.5\% of natural abundance Nd nuclei. Nd has strong dipolar hyperfine coupling \cite{Bleaney1963}, and the nuclear hyperfine splitting has been empirically shown to be linearly dependent upon the static Nd electronic moment \cite{Scheie2019,Chatterji2008}.  We can leverage this fact to estimate the root-mean-square (RMS) local static Nd moment by fitting the nuclear Schottky anomaly\cite{Scheie2019,Scheie2021magnon}, as shown in Figure 4(b). 

Although the upturn is very small, it is enough to constrain the size of the local static Nd moment. The specific heat below \textit{T} = 0.28 K fits to the equation \textit{C}(\textit{T}) = \textit{AT} + \textit{BT}$^3$ + \textit{N}(\textit{T}), where \textit{T} is temperature, \textit{A} and \textit{B} are phenomenological fitted parameters to describe the electronic specific heat, and \textit{N}(\textit{T}) is the Nd nuclear Schottky anomaly \cite{PyNuclearSchottky}. The results are shown in Figure 4(b), and yield a RMS local static moment of $\mu_\mathrm{nuc}$ = 1.01(6) $\mu_\mathrm{B}$.  This fit implies there is a static moment exceeding the 0.71(5) $\mu_\mathrm{B}$ observed from neutron diffraction, which is a lattice-averaged mean moment. This suggests static frozen Nd magnetism that does not participate in the magnetic long-range ordered pattern rather than strong moment fluctuations in the ground state. 

The magnetic contribution to heat capacity \textit{C}$_{p,\mathrm{mag}}$\textit{T}$^{-1}$ becomes Schottky-like with increasing magnetic field. Even at low applied fields, a high-temperature shoulder appears neighboring the primary magnetic ordering peak, which can also be fit alongside the peak in the zero-field data set. These shoulders can be modeled as a two-level Schottky anomaly, as shown in Figure 4(c), and a linear relationship between applied field and the Schottky gap, $\Delta$\textit{E}$_\mathrm{Schottky}$ is identified, as shown in Figure 4(d). This suggests that the gap is Zeeman-like, and the small zero-field splitting arises from a static internal magnetic field. Fitting this linear dependence to a Zeeman equation reveals a \textit{g}$_\mathrm{Zeeman}$ = 1.94(2), which is similar to the \textit{g}$_\mathrm{eff}$ = 1.86(1) value obtained from magnetization data, implying that this Schottky anomaly gap originates from a splitting of the ground state Kramers doublet. 

Furthermore, a linear extrapolation of the in-field data sets yields an estimate for the zero-field gap which is consistent with the value extracted from modeling the shoulder in the zero-field measurement, strongly suggesting a zero-field splitting on the order of 0.15(4) meV $\approx$ 1.74 K. Given the weaker interplanar antiferromagnetic coupling in this lattice that defines the transition into long-range magnetic order, the observed splitting likely originates from in-plane ferromagnetic correlations that persist far above \textit{T}$_N$.

This observation is consistent with the zero-field splitting of 0.8 meV observed in the first excited CEF level, as illustrated in Figures 3 (c) and (d). The differences in splitting (0.15 vs. 0.8 meV) for the two states suggest a mean-field origin which couples differently to each state according to their wavefunction eigenvectors---\textit{i.e.}, the ground state is mostly of $|\pm\tfrac{1}{2}\rangle$ character, whereas the first CEF excitation is mostly of $|\pm\tfrac{3}{2}\rangle$ character.  Notably, however, the magnitude of the zero-field splitting of the ground state doublet predicted using the mean-field from the CEF model is $\approx$ 0.76 meV---a value much larger than the observed 0.15 meV.  This discrepancy again likely originates due to the variance between the CEF-derived and experimentally determined $g_\mathrm{eff}$ values for the compound.

\section{Discussion}

Close to its ordering temperature, NdCd$_3$P$_3$ possesses a well-isolated $S_{\mathrm{eff}} = 1/2$ ground state in its CEF level scheme. Neutron scattering, magnetization, and heat capacity measurements reveal that the ground-state Kramers doublet is dominated by $|\pm\tfrac{1}{2}\rangle$ character; however the \textit{g}-tensor predicted by the CEF multiplet scheme solved via INS data is at variance with magnetization and neutron diffraction data.  This discrepancy likely originates from local site-symmetry lowering due to short-range bond order in the neighboring Cd$_3$P$_3$ layers \cite{arXiv2025LnCd3P3}.  

Lower point group symmetry in either an orthorhombic or monoclinic setting allows for added Stevens parameters, which can modify the mixture of states in the ground state multiplet.  Our preliminary exploration of potential solutions in this lower symmetry space with a $g_\mathrm{eff}$ matching experiments shows a large degeneracy of potential solutions that will need to be constrained by future detailed structural analysis.  So while the current, average structure CEF solution shows a planar XY-like anisotropy, an ingredient commonly associated with exotic quantum states such as spin liquids or vortex crystals \cite{Koshelev1994}, this solution will necessarily be refined once the precise local environment about the Nd sites is known. 

Regardless of the precise \textit{g}-tensor value and anisotropy, a striking observation is the apparent zero-field splitting of the ground state and first excited-state Kramers doublets at temperatures $T>>T_N$.  While dynamic coupling to a low-energy optical phonon could produce a time-averaged symmetry breaking effect, particularly if the coupling is mode-selective and sensitive to the orbital character of the CEF eigenstates, constant-energy \textit{Q}-cuts about the split CEF modes do not reveal any obvious overlap with an optical lattice mode near 6 meV.  Additionally, while this effect can drive a splitting within a select excited-state doublet \cite{arXiv2025Scheie}, it cannot drive a parallel splitting of the ground state doublet as resolved in heat capacity data.

Instead, the level-dependent splitting suggests an internal magnetic mean-field origin, wherein an effective magnetic field couples differently to each CEF level depending on their wavefunction composition. Such a field could arise from fluctuating in-plane ferromagnetic correlations above \textit{T}$_N$, which generate a quasi-static internal field that breaks time-reversal symmetry on the neutron timescale without producing long-range magnetic order.  The ordering temperature \textit{T}$_N$ = 0.34 K, marks the onset of commensurate \textit{A}-type antiferromagnetic order; however the dominant energy scale is expected to the ferromagnetic coupling of Nd moments within each triangular lattice plane that seemingly persist at $T>>T_N$.  This occurs despite previous assessment of the mean field $\Theta_{CW}\approx T_N$ from magnetization measurements and is indicative of magnetic frustration effects that were previously hidden.

Comparison of the ordered moment from neutron diffraction $\langle \mu \rangle$ = 0.71(5) $\mu_\mathrm{B}$, the nuclear Schottky moment $\mu_\mathrm{nuc}$ = 1.01(6) $\mu_\mathrm{B}$, and the saturation moment from high-field magnetization $\mu_\mathrm{sat}$ = 0.934(3) $\mu_\mathrm{B}$ reveals a reduced moment in the antiferromagnetically ordered state. While a sizable fraction of the Nd moments are static, they are not all fully participating in long-range magnetic order. The nuclear Schottky-derived moment implies that static local moments exist even in the absence of long-range coherence, and this quasi-static or glassy magnetism is further supported by heat capacity data, which shows that the magnetic entropy is only gradually released, reaching Rln(2) near \textit{T} = 10 K \cite{Chamorro2023}. This is well above the ordering temperature \textit{T}$_N$ = 0.34 K. These findings imply substantial short-range correlations and possible time-reversal symmetry breaking fluctuations that persist over a broad temperature window, highlighting the importance and need for further low-temperature probes such as $\mu$SR or single-crystal neutron diffraction.

Finally, it is instructive to compare NdCd$_3$P$_3$ to other triangular-lattice compounds containing Nd$^{3+}$ ions, where differing crystal fields and exchange interactions lead to a range of magnetic ground states. A natural comparison is isostructural NdZn$_3$P$_3$. Previous reports have indicated strong anisotropy, with dominant Ising character along the \textit{c}-axis, resulting in an Ising-like order \cite{Kabeya2020} that occurs at \textit{T}$_N$ = 0.63 K. Rln(2) entropy is not fully recovered until \textit{T} $\approx$ 4 K, implying the presence of magnetic interactions far above \textit{T}$_N$ in that system as well. In both Zn- and Cd-based compounds, Nd$^{3+}$ should experience similar chemical environments, \textit{i.e.}, both contain NdP$_6$ octahedra; however differences in anisotropies may arise due to differing bond distances of 2.943(1) $\mathrm{\AA}$ in NdCd$_3$P$_3$ versus 2.902(1) $\mathrm{\AA}$ in NdZn$_3$P$_3$ \cite{Nientiedt1999}. Perhaps more importantly, differences in anisotropy and CEF splitting may also be expected due to local ordering effects in the interconnecting cadmium phosphide layer in the Cd-based variant \cite{arXiv2025LnCd3P3}. Future inelastic neutron scattering measurements are required to assess whether the CEF level scheme of NdZn$_3$P$_3$ differs significantly from that of NdCd$_3$P$_3$ and whether CEF doublets are similarly split at \textit{T} $>$ \textit{T}$_N$.

\section{Conclusion}

 NdCd$_3$P$_3$ is a triangular-lattice antiferromagnet hosting an $S_{\mathrm{eff}}$ = 1/2 ground state that harbors signs of magnetic frustration via short-range magnetic correlation effects observed at temperatures $T>>T_N$. This compound undergoes a transition to \textit{A}-type antiferromagnetic ordering below $T_N = 0.34$ K; however zero-field splitting is observed in both the ground-state and the first excited CEF doublets at temperatures of at least 17$T_N$.  This suggests the presence of broken time-reversal symmetry acting well above the ordering temperature. Notable variance between the ordered magnetic moment and the field-polarized magnetization also suggest an unusual thermally extended regime where short-range order and fluctuation effects persist across the triangular lattice network of Nd$^{3+}$ moments.

\section*{Acknowledgments}
This work was supported by the US Department of Energy (DOE), Office of Basic Energy Sciences, Division of Materials Sciences and Engineering under Grant No. DE-SC0017752. J.R.C. acknowledges additional support through the NSF MPS-Ascend Postdoctoral Fellowship (DMR-2137580). The work by A.S. is supported by the Quantum Science Center (QSC), a National Quantum Information Science Research Center of the U.S. Department of Energy (DOE). The MRL Shared Experimental Facilities are supported by the MRSEC Program of the NSF under Award No. DMR 2308708; a member of the NSF-funded Materials Research Facilities Network (www.mrfn.org). This work used facilities supported via the UC Santa Barbara NSF Quantum Foundry funded via the Q-AMASE-i program under award DMR-1906325. A portion of the research used resources at the Spallation Neutron Source, a DOE Office of Science User Facility operated by the Oak Ridge National Laboratory. The beam time was allocated to SEQUOIA spectrometer on Proposal No. IPTS-30819.1. This work is also based on experiments performed at the Swiss Spallation Neutron Source SINQ, Paul Scherrer Institute, Villigen, Switzerland.

\bibliography{biblio}

\end{document}